\documentclass[aps, 10pt, prd,
               notitlepage, twocolumn, superscriptaddress,
               nofootinbib, floatfix,
               longbibliography]{revtex4-1}

\usepackage{amsmath}
\usepackage{amssymb}
\usepackage{amsfonts}
\usepackage[utf8]{inputenc}
\usepackage[T1]{fontenc}
\usepackage{mathrsfs}
\usepackage{anyfontsize}
\usepackage{aas_macros}
\usepackage{float}
\usepackage{placeins}

\usepackage[linktocpage,breaklinks]{hyperref}
\usepackage[usenames,dvipsnames]{xcolor}

\usepackage{txfonts}
\usepackage{tensor}
\usepackage{bm}

\usepackage{graphicx}
\usepackage{epsfig}
\usepackage{epstopdf}
\usepackage{dsfont}

\usepackage{natbib}
\usepackage{hyperref}
\hypersetup{colorlinks=true,
            citecolor=MidnightBlue,
            linkcolor=MidnightBlue,
            urlcolor=MidnightBlue}

\newcommand{\cC}{\mathscr{C}}
\newcommand{\cR}{\mathscr{R}}
\newcommand{\msun}{\textrm{M}_{\odot}}
\newcommand{\bfi}[1]{\textit{#1$-$}}

\newcommand{\bi}{\bar{I}}
\newcommand{\bl}{\bar{\lambda}}
\newcommand{\bq}{\bar{Q}}

\newcommand{\bx}{\bar{\xi}}

\newcommand{\dd}{{\rm d}}
\newcommand{\GR}{{\mbox{\tiny GR}}}
\newcommand{\crit}{{\mbox{\tiny crit}}}
\newcommand{\CSN}{{\mbox{\tiny CS,N}}}
\newcommand{\CS}{{\mbox{\tiny CS}}}
\newcommand{\N}{{\mbox{\tiny N}}}
\newcommand{\pa}{{1.3381}}

\newcommand{\id}{PSR~J0030+0451}
\newcommand{\idd}{\id\,}
\newcommand{\iddp}{PSR~J0737-3039A}
\newcommand{\iddps}{J0737-3039}

\AtBeginDocument{%
    \newwrite\bibnotes
    \def\bibnotesext{Notes.bib}
    \immediate\openout\bibnotes=\jobname\bibnotesext
    \immediate\write\bibnotes{@CONTROL{REVTEX41Control}}
    \immediate\write\bibnotes{@CONTROL{%
    apsrev41Control,author="08",editor="1",pages="1",title="0",year="1"}}
    \if@filesw
    \immediate\write\@auxout{\string\citation{apsrev41Control}}%
    \fi
}

\begin{document}

\title{Astrophysical and theoretical physics implications from multimessenger
neutron star observations}

\author{Hector O. Silva}
\affiliation{Max-Planck-Institut f\"ur Gravitationsphysik (Albert-Einstein-Institut),
Am M\"uhlenberg 1, D-14476 Potsdam, Germany}
\affiliation{Department of Physics and Illinois Center for Advanced Studies of the Universe,
University of Illinois at Urbana-Champaign,
Urbana, Illinois 61801, USA}

\author{A. Miguel Holgado}
\affiliation{Department of Physics and McWilliams Center for Cosmology,
Carnegie Mellon University,
Pittsburgh, Pennsylvania, 15213, USA}
\affiliation{Department of Astronomy,
University of Illinois at Urbana-Champaign,
Urbana, Illinois 61801, USA}
\affiliation{National Center for Supercomputing Applications, University of
Illinois at Urbana-Champaign, Urbana, Illinois 61801, USA}

\author{Alejandro C\'ardenas-Avenda\~{n}o}
\affiliation{Department of Physics and Illinois Center for Advanced Studies of the Universe,
University of Illinois at Urbana-Champaign,
Urbana, Illinois 61801, USA}
\affiliation{Programa de Matem\'atica, Fundaci\'on Universitaria Konrad Lorenz,
110231 Bogot\'a, Colombia}

\author{Nicol\'as Yunes}
\affiliation{Department of Physics and Illinois Center for Advanced Studies of the Universe,
University of Illinois at Urbana-Champaign,
Urbana, Illinois 61801, USA}

\begin{abstract}
The Neutron Star Interior Composition Explorer (NICER) recently
measured the mass and equatorial radius of the isolated neutron star \id.
We use these measurements to infer the moment of inertia, the quadrupole
moment, and the surface eccentricity of an isolated neutron star for the first
time, using relations between these quantities that are insensitive to the
unknown equation of state of supranuclear matter.
We also use these results to forecast the moment of inertia of neutron star $A$
in the double pulsar binary \iddps, a quantity anticipated to be directly
measured in the coming decade with radio observations.
Combining this information with the measurement of the tidal Love number with
LIGO/Virgo observations, we propose and implement the first theory-agnostic and
equation-of-state-insensitive test of general relativity.
Specializing these constraints to a particular modified theory, we find that
consistency with general relativity places the most stringent constraint on
gravitational parity violation to date, surpassing all other previously
reported bounds by 7 orders of magnitude and opens the path for a future test
of general relativity with multimessenger neutron star observations.
\end{abstract}

\maketitle

\bfi{Introduction.}
%
Neutron stars are some of the most extreme objects in nature.
Their mass (typically around $1.4\,\msun$) combined with their small radius
(between $10-14$ km) result in interior densities that can exceed nuclear
saturation density ($\rho \geqslant 2.8 \times 10^{14}$ g/cm$^{3}$), above which
exotic states of matter can arise~\cite{Baym:2017whm}.
Neutron stars are, next to black holes, the strongest gravitational field
sources known, with typical gravitational potentials that are 5 orders of
magnitude larger than that of the Sun.
These properties make neutron stars outstanding laboratories to study both
matter and gravity in situations out of reach in terrestrial and Solar System
experiments.

Our current poor understanding of the supranuclear equation of state
translates, via the equations of stellar equilibrium, to a large variability on
observable properties of neutron stars, such as their masses and
radii~\cite{Ozel:2016oaf}.
This variability increases if one lifts the assumption that Einstein's theory
of general relativity is valid in the strong-gravity environment of neutron
star interiors~\cite{Berti:2015itd}. Modifications to general relativity
generically predict new equations of stellar equilibrium, which, when combined
with uncertainties on the nuclear equation of state, jeopardize attempts to
test Einstein's theory with isolated, neutron star observations.

One possibility to circumvent this issue is to explore whether relations
between neutron-star observables that are insensitive to either (or both) the
equation of state and the gravitational theory exist.
Fortunately, they do. For instance, when properly nondimensionalized, the
moment of inertia ($I$), the rotational quadrupole moment $(Q)$ and the tidal
Love number ($\lambda$) of neutron stars show a remarkable degree of
equation-of-state insensitivity, at a level below 1\%~\cite{Yagi:2013bca,Yagi:2013awa}.
These ``I-Love-Q'' relations also exist in some modified theories of
gravity, although they are different from their general relativity
counterparts~\cite{Doneva:2017jop}.

We here combine the first measurements~\cite{Riley:2019yda,Miller:2019cac} by
NICER~\cite{GendreuNICER} of \textit{both} the mass ($M$) and equatorial radius
($R_{\rm e}$) of the isolated pulsar
\id~\cite{Lommen:2000yt,Arzoumanian:2017puf} with known equation-of-state insensitive
relations involving the compactness
$\cC = G M / (R_{\rm e} c^2)$
~(see, for instance Refs.~\cite{Lattimer:2015nhk,Maselli:2013mva,Baubock:2013gna,Breu:2016ufb})
to infer a number of astrophysical and
theoretical physics consequences.
Before doing so, let us explain how these relations are obtained.

\bfi{Quasiuniversal relations.}
%
Neutron stars can have short rotation periods of the order of milliseconds, so
their surfaces are oblate instead of spherical.
The inclusion of this effect is of critical importance to accurately model the
thermal x-ray waveform that NICER observes, since the x-rays are produced by
hotspots at the star's surface~\cite{Morsink:2007tv,Bogdanov:2019qjb}.
The canonical approach to model relativistic rotating stars was developed in
the 1970s~\cite{Hartle:1967he,Hartle:1968si}.
In this approach, the star's rotation is treated as a small perturbation
$\varepsilon = f / f_0 \ll 1$, involving the star's rotation frequency $f$ and
its characteristic mass-shedding frequency $f_0 = (G M / R_{\rm e}^3)^{1/2} / (2\pi)$.
Rotating stars are then found by perturbing in $\varepsilon$ an otherwise
nonrotating star, which can be obtained by solving the
Tolman-Oppenheimer-Volkoff equations~\cite{Shapiro:1983du}.
This slow-rotation approximation is well justified for most neutron stars with
astrophysically relevant spins. Even for a prototypical millisecond pulsar with
$f = 700$ Hz, $M = 1.4\,\msun$ and $R_{\rm e} = 11$ km, one has $\varepsilon = 0.37$.
In the case of \id, its rotation frequency is known to be $f_{\star} = 205.53$
Hz~\cite{Lommen:2000yt,Arzoumanian:2017puf}, so $\varepsilon_{\star} = 0.14$,
when one uses the best-fit $M$ and $R_{\rm e}$ values obtained by
NICER~\cite{Riley:2019yda,Miller:2019cac}.
Henceforth, a ``$\star$'' indicates observables associated with \idd.

Using this technique, we numerically calculated over a thousand neutron
star solutions to order $\varepsilon^2$ in this perturbative scheme, using a
broad set of 46 different equations of state~\cite{Read:2008iy,Kumar:2019xgp},
as detailed in the Supplemental Material (SM).
From these solutions, we then numerically computed the moment of inertia $I$,
the rotational quadrupole moment $Q$, the surface eccentricity $e$, and
the electric-type, $\ell = 2$, tidal Love number $\lambda$,
which is the dominant parameter in the description of
tidal effects in the late inspiral of neutron star
binaries~\cite{Mora:2002gf,Flanagan:2007ix,Hinderer:2007mb}.
We nondimensionalized these quantities through division by the appropriate
factors of $M$ and dimensionless spin
$\chi = \left( 2 \pi f_{0} \right) G \bar{I} M / c^{3}$,
namely:
$\bi = c^4 I / (G^2 M^3)$,
$\bar{Q}=-c^{4}Q/\left(G^{2}M^{3}\chi^{2}\right)$
and
$\bl =  c^{10} \lambda / (G M)^{5}$.
The surface eccentricity $e$ is dimensionless by definition,
given in terms of the star's equatorial $R_{\rm e}$ and polar $R_{\rm p}$
radii as $e = [(R_{\rm e} / R_{\rm p})^2 - 1]^{1/2}$~\cite{Baubock:2013gna}.
The relations between these nondimensional quantities
are strongly insensitive to the equation of state.
Because of the small value of $\varepsilon_{\star}$ we can neglect
higher order in spin corrections in this expression.

The first step in using the approximately universal relations on NICER's first
observation is to derive equation-of-state-insensitive relations between the
observables $\{\bi$, $\bar{Q}$, $\bl$, $e\}$, with respect to the compactness
$\cC$.
Details of these ``$\cC$ relations'' are given in the SM.
Our plan of attack is then clear: use the publicly available
Markov~Chain~Monte~Carlo~(MCMC) $M$-$R_{\rm e}$
samples~\cite{miller_m_c_2019_3473466,riley_thomas_e_2020_3707821} for the
best-fit model inferred by two independent
analysis~\cite{Miller:2019cac,Riley:2019yda} of the NICER
data~\cite{Bogdanov:2019ixe}.
Although each group modeled the surface hotspots differently and used different sampling methods, their results are consistent with each other.
Here we use the results for the three-hotspot model inferred by Miller et al.~\cite{Miller:2019cac}
and the favored single temperature, two-hotpot ST+PST model from Riley et
al.~\cite{Riley:2019yda} to obtain a posterior distribution for the
compactness, and then use the approximately universal relations to infer other
astrophysical quantities.
We detail this procedure next.

\bfi{Astrophysical implications.}
%
We begin by constructing a posterior distribution $P(\cC|\textrm{NICER})$ for
the compactness $\cC$ of \id, using the MCMC
chains~\cite{miller_m_c_2019_3473466,riley_thomas_e_2020_3707821}.
With this posterior in hand, we then use the $\cC$ relations to inferred
posterior distributions for $\{\bi$, $\bl$, $\bar{Q}$, $e\}$.

The implementation of such an inference procedure requires a particular
scheme, and we here follow a proposal that accounts for the approximately
universal nature of the relations~\cite{Kumar:2019xgp}.
In this scheme, the maximum relative error of each fitting function defines the
half width of the 90\% credible interval of a Gaussian distribution centered at
each fitted value.
The posterior distribution for each dimensionless quantity is then calculated
using the corresponding $\cC$ relation and the posterior distribution of the
compactness, after marginalizing over the latter.
From these posteriors and using the same procedure described above, we can also
construct posteriors for the dimensionful versions of these quantities by a
change of variables, marginalize over the nuisance variables mass $M$ and
radius $R_{\rm e}$, and then do a final rescaling of the posterior by
$\varepsilon$ ($=0.14$) for the surface eccentricity $e$ and by $\varepsilon^2$
for the rotational quadrupole moment $Q$.
We refer to the SM for details.

The resulting mean and $1\sigma$ intervals of these parameters
(both the nondimensionalized and the dimensionful versions) are shown
in~Table~\ref{tab:gr_infer}; see the SM section for plots of the inferred
posteriors.
The reported confidence intervals in all of these quantities account for both
the approximate nature of the universal relations and the uncertainties in
NICER's observation.
These results are the \emph{first inferences on the moment
of inertia, the surface eccentricity, the Love number and the quadrupole moment
of an isolated neutron star.}

\begin{table}[t]
\begin{tabular}{ l c c }
\hline
\hline
Parameter  & Miller et al.  & Riley et al.  \\
\hline
$\bar{I}_{\star}$ \,\,($10$)  & $1.31^{+0.13}_{-0.11}$ & $1.42^{+0.26}_{-0.19}$ \\
$\bar{\lambda}_{\star}$  \,($10^2$) &  $4.97^{+1.92}_{-1.28}$  & $6.75^{+5.52}_{-2.69}$  \\
$\bar{Q}_{\star}$   & $5.92^{+0.73}_{-0.61}$ & $6.50^{+1.38}_{-1.03}$  \\
\hline
$I_{\star}$ \,\,($10^{45} $  g cm$^2$) &  $1.71^{+0.64}_{-0.48}$  & $1.42^{+0.81}_{-0.53}$ \\
$Q_{\star}$ ($10^{43}$ g cm$^2$)  &  $1.49^{+0.63}_{-0.45}$ & $1.27^{+0.74}_{-0.49}$ \\
$e_{\star} \,\,\, (10^{-1})$  &  $1.56^{+0.25}_{-0.21}$  &  $1.58^{+0.29}_{-0.28}$ \\
\hline
\hline
\end{tabular}
\caption{Inferred properties of \idd using equation-of-state-insensitive
relations combined with the MCMC samples by Miller et al.~\cite{miller_m_c_2019_3473466}
and Riley et al.~\cite{riley_thomas_e_2020_3707821}.
We report the values within 1 standard deviation from the
mean, representing approximately 68\% confidence intervals.
These values are the first inferences of the moment of inertia, the
eccentricity, the Love number, and the quadrupole moment of an isolated neutron
star.
}
\label{tab:gr_infer}
\end{table}

We can also use NICER's observation combined with the I-C relation to estimate
the moment of inertia of \iddp~($I_\pa$), where the subscript refers to this
pulsar's measured mass of $M = (1.3381 \pm 0.0007)\,\msun$~\cite{Kramer:2006nb}.
The double pulsar J0737-3039 is expected to provide the first direct neutron star
measurement of the moment of inertia~\cite{Kramer:2009zza}.
This system is the only double-pulsar observed to date, which makes it
an unique laboratory for binary stellar
astrophysics~\cite{Stairs:2006na,Ferdman:2013xia}.
Moreover, an accurate measurement of $I_\pa$ in combination with its known mass is
expected to strongly constrain the nuclear equation of state
around once and twice nuclear saturation density~\cite{Lattimer:2004nj}.

To predict the moment of inertia of \iddp\, from NICER's observation of PSR J0030+0451, we
first need to obtain an estimate for the compactness ${\cC}_\pa$ of \iddp.
This can be approximated by the substitution
$\{M, R_{\rm e}\} \mapsto \{M_0\!=\!1.3381\,\msun, \, R_{\rm e}\}$
at each MCMC sample~\cite{miller_m_c_2019_3473466} and then computing $\cC_{M_0}$.
This yields an approximation to the distribution of compactness
for a system with mass $M_0$, which is assumed known and identical to \id.
This procedure is only justified as long as $M_0$ is very close to
$M_{\star}$, as in the case of \iddp, whose inferred mass ($M_{0} = 1.3381^{+0.0007}_{-0.0007} \,\msun$)~\cite{Kramer:2006nb}
is within the $1\sigma$ credible interval of both NICER's mass inference
($M_{\star} = 1.34^{+0.15}_{-0.16}\,\msun$~\cite{Riley:2019yda} and $M_{\star} = 1.44^{+0.15}_{-0.14}\,\msun$~\cite{Miller:2019cac}).

\begin{figure}[t]
\includegraphics[width=\columnwidth]{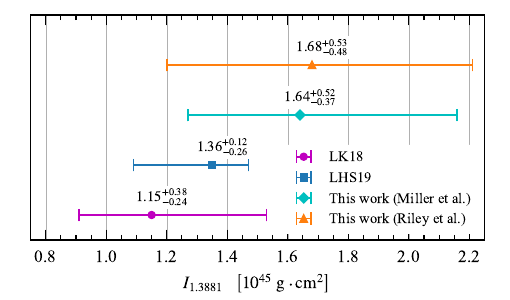}
\caption{
Predictions for the moment of inertia of \iddp. We compare our predicted
$I_\pa$ using both the MCMC samples from
Miller~et~al.~\cite{miller_m_c_2019_3473466} and
Riley~et~al.~\cite{riley_thomas_e_2020_3707821} against:
(i) Landry and Kumar~\cite{Landry:2018jyg} (``LK18''), which used binary
Love~\cite{Yagi:2015pkc} and I-Love relations with the tidal-deformability
constraints from binary neutron-star merger GW170817~\cite{Abbott:2018wiz}, and
(ii) Lim et al.~\cite{Lim:2018xne} (``LHS19'') which carried out Bayesian
modeling of a number of equations of state.
The larger moment of inertia that we predict is due to the larger radii favored
by an $M \approx 1.4$~M$_{\odot}$ neutron star by NICER's observation relative
what is inferred by the two other methods, as $I \sim M R^2_{\rm e}$.
}
\label{fig:i_j0737}
\end{figure}

With an estimate of the compactness of \iddp, we can now obtain a prediction
for \id's moment of inertia repeating the procedure applied to \id.
Figure~\ref{fig:i_j0737} shows our result using both NICER MCMC samples;
$I^{\textrm{Miller et al.}}_\pa = 1.64^{+0.52}_{-0.37} \times 10^{45}$ g cm$^{2}$,
and
$I^{\textrm{Riley et al.}}_\pa = 1.68^{+0.53}_{-0.48} \times 10^{45}$ g cm$^{2}$,
together with two other independent predictions~\cite{Landry:2018jyg,Lim:2018xne}.
All predictions are consistent with one another.
The anticipated future independent measurement of $I_\pa$ from continued radio
timing of \iddp~will provide another test for nuclear theory and enable an
``I-Love test'' of gravity, the latter of which we define next.

\bfi{Theoretical physics implications.}
%
The combination of the inference of $I$ with NICER data described above, and
the independent measurement of $\lambda$~\cite{Abbott:2018wiz} by the
LIGO/Virgo Collaboration from the binary neutron-star merger
GW170817~\cite{TheLIGOScientific:2017qsa}, allows for the first
implementation of an I-Love test~\cite{Yagi:2013bca}.
This test would be \emph{the first multimessenger test of general relativity
with neutron star observables}.

\begin{figure}[t]
\includegraphics[width=\columnwidth]{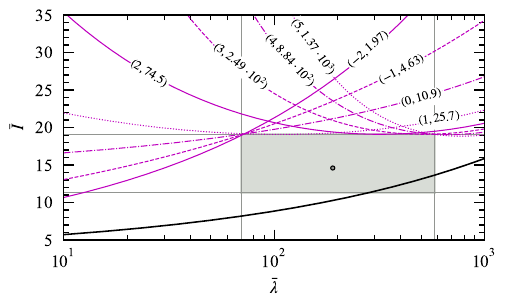}
\caption{Multimessenger test of general relativity using the parametrized
I-Love relation.
The vertical (horizontal) lines delimit the 90\% confidence region (shaded) for
$\bl_{1.4}$~\cite{Abbott:2018exr} ($\bi_{1.4}$, this work), while the circle
marks the median ($190$, $14.6$).
The solid black line corresponds to the I-Love relation in general relativity
[Eq.~\eqref{eq:ilove}] and is consistent with the inferred values of
$\bi_{1.4}$, $\bl_{1.4}$ at 90\% confidence.
Starting from $b=-2$ and moving clockwise, we show the parametrized I-Love
curves $(b,\, \beta_{\crit})$, where $b \in [-2, 5]$ and
$\beta_{\crit}$ is the critical value of $\beta$ above which the parametrized
I-Love relation [Eq.~\eqref{eq:ilove_param}] fails to pass by the 90\%
confidence region in the plane.
Here we used the value of $\bi_{1.4}$ inferred using the results by Miller et
al.~\cite{Miller:2019cac,miller_m_c_2019_3473466}. We found similar results
using the results by Riley et al.~\cite{Riley:2019yda,riley_thomas_e_2020_3707821}
(See SM).
}
\label{fig:pilove}
\end{figure}

The idea of an I-Love test is as follows~\cite{Yagi:2013bca,Yagi:2013awa} (see
Fig.~\ref{fig:pilove}). Consider two independent inferences of $\bi_{1.4}$
and $\bl_{1.4}$ for a $1.4 \, \msun$ neutron star.
In the ($\bi$, $\bl$) plane, these measurements yield a 90\% confidence error box.
If the I-Love relation in general relativity, including its small
equation-of-state variability, \emph{does not} pass through this error box,
then there is evidence for a violation of Einstein's theory, regardless of the
underlying  equation of state.
Moreover, if any theory of gravity predicts an I-Love curve that also does not
pass through this error box for a given value of its coupling constants, then
the I-Love test places a constraint on the couplings of this theory, which is
also independent of the equation of state.

Such a test, however, requires the inference of the tidal deformability and the
moment of inertia of a neutron star of the \emph{same} mass.
The LIGO/Virgo Collaboration used gravitational wave data to infer the tidal
deformability of a $1.4 \, \msun$ neutron star to be
$\bl_{1.4} = 190^{+390}_{-120}$
at 90\% confidence~\cite{Abbott:2018exr}, obtained under the assumptions
that the binary components were described by the same equation of state and
were slowly spinning.
We can use NICER's data to infer the moment of inertia of a $1.4 \, \msun$ neutron star
with the same techniques we used to predict the moment of inertia of
\iddp.
For concreteness, we use the results from Miller et al.~\cite{Miller:2019cac,miller_m_c_2019_3473466},
but we verified (see the SM for detail) that our conclusions are essentially the same
had we used the results from Riley et al.~\cite{Riley:2019yda,riley_thomas_e_2020_3707821}.
We find that
$\cC_{1.4} = 0.159_{-0.022}^{+0.025}$
and
$\bi_{1.4} =14.6_{-3.3}^{+4.5}$
at 90\% confidence.
An important underlying assumption behind both inferences is that
general relativity is the correct theory of gravity.
The rationale behind this test is detailed in the SM.

Since carrying out such a test on a theory-by-theory basis would, in general, be
complicated and time consuming, we here develop and implement a useful
\emph{parametrization} of the I-Love test.
From Newtonian gravity, we know that $\bi$ scales with $\cC^{-2}$, whereas
$\bl$ scales with $\cC^{-5}$. Therefore, $\bi = C_{\bi\bl} \bl^{2/5}$, with
$C_{\bi\bl} \approx 0.52$ a constant that depends on the equation of state very
weakly~\cite{Yagi:2013awa}.
This calculation can be extended, systematically, in a \emph{post-Minkowskian
expansion}, i.e.,~an expansion in powers of $\cC \ll 1$~\cite{Chan:2014tva}.
The outcome is that both $\bi$ and $\bl$ can be written as a power series in
$\cC$ and then be combined (as just described in the Newtonian limit) to obtain
$\bi = \bi(\bl)$.
The resulting I-Love relation has the same degree of equation-of-state
independence as the original I-Love relation~\cite{Yagi:2013bca}.
For our neutron star catalog, a parametrization in general relativity of the
form
\begin{equation}
\bi_{\GR} = \bl^{2/5}
\left(c_0 + c_1 \bl^{-1/5} + c_2 \bl^{-2/5}\right)\,,
\label{eq:ilove}
\end{equation}
with $c_0 = 0.584$, $c_1 = 0.980$, $c_2 = 2.695$, is sufficient
to reproduce our numerical data with mean relative error
$\langle \epsilon^{\bi} \rangle \leqslant 2 \times 10^{-3}$.
The prefactor $\bl^{2/5}$ is the Newtonian result, while the powers of
$\bl^{-1/5}$ inside parenthesis are relativistic (post-Minkowskian) corrections
because $\bl^{-1/5} \propto \cC \lesssim 0.2$.
Given this, we then propose a minimal deformation of the Einsteinian parametrization
in Eq.~\eqref{eq:ilove} of the form
\begin{equation}
\bi_{\rm p} = \bi_{\GR} + \beta \; \bl^{-b/5}\,,
\qquad \beta \in \mathds{R}_{+}\,,
\quad  b \in \mathds{Z}\,,
\label{eq:ilove_param}
\end{equation}
where $\beta$ and $b$ are deformation parameters that control the magnitude and
type of the deviations from general relativity in the I-Love relation
respectively. Such a parametrization is similar to that successfully used
in gravitational-wave tests of general relativity by the LIGO/Virgo
Collaboration, the parametrized post-Einsteinian
framework~\cite{Yunes:2009ke}.

We performed such a test of general relativity through the procedure described
earlier.
First, we see that the I-Love relation in general relativity does indeed
pass this null-test and it is consistent with the error box.
Second, we considered $b \in [-2, 5]$, where the lower limit is set by
requiring no deviations at the Newtonian level and the upper limit is set for
simplicity.
We then fixed $b$ and calculated what the corresponding value of
$\beta = \beta_{\crit}$ is, above which the parametrized I-Love
relation~\eqref{eq:ilove_param} would be in tension with the inferred
$(\bi_{1.4},\, \bl_{1.4})$ region at 90\% confidence.
Our results are summarized in Fig.~\ref{fig:pilove}, where the numbers
in parenthesis correspond to $(b,\,\beta_{\crit})$.
We stress that our results for $b \leqslant 0$ are of course dependent on the
posterior used for $\bl_{1.4}$. If one treated the tidal deformabilities as independent
free parameters in the waveform model~\cite{TheLIGOScientific:2017qsa}, then
the  $\bl_{1.4}$ posterior would not have a lower limit, allowing all curves
with $b \leqslant 0$ to be consistent with both observations.

With these theory-agnostic constraints in hand, we can now map them to specific
theories and place constraints on their coupling parameters.
As an example, let us consider dynamical Chern-Simons gravity, a theory that
modifies general relativity by introducing gravitational
parity violation~\cite{Jackiw:2003pm}.
This theory has found applications to several open problems in cosmology, such
as the matter-antimatter asymmetry and
leptogenesis~\cite{Weinberg:2008mc,GarciaBellido:2003wd,Alexander:2004xd,Alexander:2004us}.
It also arises in several approaches to quantum gravity, such as string
theory~\cite{Adak:2008yg} and loop quantum
gravity~\cite{Ashtekar:1988sw,Mercuri:2009zt,Smolin:1994qb}.
Mathematical well-posedness requires the theory to be treated as an effective
field theory~\cite{Delsate:2014hba}. In this formalism, one works in a
small-coupling approximation
$\zeta \equiv 16 \pi \alpha^2 \cR^{-4} \ll 1$,
where $\cR = [{c^{2} R_{\rm e}^3 / (G M)}]^{1/2}$ is the curvature length scale
associated with a neutron star (in our case), and where $\alpha$ is a coupling
constant with units of length squared, such that $\zeta$ is dimensionless.
This theory modifies Einstein's only when gravity is strong, and thus, it
passes all Solar System constraints, being only extremely weakly constrained by
Gravity Probe B and the LAGEOS satellites, and table-top experiments,
to
${\alpha}^{1/2} \leqslant 10^{8} \, {\rm km}$ \cite{Alexander:2009tp,Yagi:2012ya,Nakamura:2018yaw}.
This theory has also evaded gravitational-wave tests~\cite{Nair:2019iur},
making it a key target to test the constraining power of our new I-Love test.

Let us now map the theory-agnostic deformation of the I-Love relations in
Eq.~\eqref{eq:ilove_param} to dynamical Chern-Simons gravity, though this
methodology could be applied to other theories as well.
As we discuss in the SM, the I-Love relation in this theory can be
described by Eq.~\eqref{eq:ilove_param} with
$b_{\CS}=4$
and
$\beta_{\CS} = 6.15 \times 10^{-2} \bx$, where $\bx = 16 \pi \alpha^{2} / M^4$.
We can now use our theory-agnostic constraints on $\beta$ to place a constraint
on $\alpha$, the coupling constant of dynamical Chern-Simons gravity.
Using the constraint on $\beta$ when $b = 4$, namely
$\beta_{\crit} \leqslant 8.84 \times 10^{2}$,
and applying the mapping, yields
$\beta_{\CS}  =  6.15 \times 10^{-2} \, \bx \leqslant 8.84 \times 10^{2}$,
or simply
\begin{equation}
{\alpha}^{1/2} \leqslant 8.5 \,\textrm{km}\,,
\label{eq:alpha_bound}
\end{equation}
at 90\% credibility, \emph{if the theory is to be consistent with
the observational bounds on $\bi_{1.4}$ and $\bl_{1.4}$}.
Using the mean value $\cC_{1.4} = 0.159$, which implies the mean equatorial
radius $R_{1.4} = 13.0$~km, we also find that $\zeta \leq 0.23$ when using
Eq.~\eqref{eq:alpha_bound}, implying that the small-coupling approximation is
indeed satisfied.
This bound is \emph{7 orders of magnitude stronger than any previous
constraints and it is unlikely to be improved upon
with foreseeable gravitational-wave observations~\cite{Alexander:2017jmt}}.

\bfi{Conclusions and outlook.}
%
NICER's observation of \idd allows the extraction of new astrophysical and
theoretical physics inferences when one uses equation-of-state-insensitive
relations.
We have here shown the first inferences of the moment of inertia, the
quadrupole moment, the surface eccentricity and the Love number of an isolated
neutron star. We have also been able to perform the first theory-agnostic and
equation-of-state independent test of general relativity by combining NICER and
LIGO/Virgo's observations.
This test, in turn, was leveraged to produce the most stringent constraint on
gravitational parity violation, improving previous bounds by 7 orders of
magnitude.
This robust methodology can be applied to future multimessenger observations of
neutron stars with NICER and gravitational wave observatories, with important
implications to nuclear astrophysics and theoretical physics.

\textit{Acknowledgments.}~We thank Toral Gupta, Fred Lamb, Philippe Landry, and
Helvi Witek for various discussions.
We also thank Cole Miller, Sharon Morsink and Kent Yagi for suggestions that
improved this work.
We thank the NICER Collaboration for making~\cite{miller_m_c_2019_3473466,riley_thomas_e_2020_3707821}
publicly available.
H.O.S, A.C.-A. and N.Y. are supported by NASA Grants
Nos.~NNX16AB98G,~80NSSC17M0041,~80NSSC18K1352
and NSF Award No.~1759615.
A.C.-A. also acknowledges funding from the Fundaci\'on Universitaria
Konrad Lorenz (Project 5INV1).
A.M.H. was supported by the DOE NNSA Stewardship Science Graduate
Fellowship under Grant No.~DE-NA0003864.

\bibliography{biblio}

\clearpage
\newpage

\begin{center}
    \color{Sepia}{{-- Supplemental Material --}}
\end{center}

\appendix

\bfi{Equation of state catalog.}
%
We consider the large catalog of nuclear
equations of state of~\cite{Kumar:2019xgp}, supplemented by the non-duplicates
from~\cite{Read:2008iy}.
To this set of 85 equations of state, we impose the following observational
consistency criteria.
First, the equation of state must be consistent with the 90\% confidence region
of the two-dimensional marginalized posterior ($M$, $R_{\rm e}$) reported by the
LIGO/Virgo collaboration~\cite{Abbott:2018exr} for each component of the binary
neutron-star merger GW170817.
Second, the equation of state must be consistent with the 90\% confidence
region of the two-dimensional marginalized posterior  ($M$, $R_{\rm e}$) reported
by NICER~\cite{Miller:2019cac,Riley:2019yda} for \id.
Third, the equation of state must allow for neutron stars with masses above
$M_{\rm max} \geqslant 1.96\,\msun$,
which corresponds to the lower-limit estimate of the most massive neutron star
known, the millisecond pulsar~J0740+6620, at 95.4\% confidence
level~\cite{Cromartie:2019kug}.
In total, this yields the following 46 equations of state:
ALF2, APR3, APR4, BCPM BSP, BSR2, BSR2Y, BSk20, BSk21, BSk22, BSk23, BSk24,
BSk25, BSk26, DD2, DD2Y, DDHd, DDME2, DDME2Y, ENG, FSUGarnet, G3, GNH3, IOPB,
K255, KDE0v1, MPA1, Model1, Rs, SINPA, SK272, SKOp, SKa, SKb, SLY2, SLY230a,
SLY4, SLY9, SLy, SkI2, SkI3, SkI4, SkI6, SkMP, WFF1 and
WFF2~\cite{Read:2008iy,Kumar:2019xgp}.

Our numerical code to calculate neutron stars (in general relativity and
dynamical Chern-Simons gravity) uses the piecewise-polytrope approximation to
model these equations of state~\cite{Read:2008iy}.
In this approximation, the relationship between pressure ($p_i$) and baryonic
mass density ($\rho_i$) is given by $p_i = K_i\, \rho^{\Gamma_i}$, where $K_i$
is the polytropic index and $\Gamma_i$ the adiabatic index, on a sequence of
baryonic mass density intervals $i$.
For the high-baryonic mass densities describing the neutron star core
($\rho_{\rm core} \gtrsim 1.7 \times 10^{14}$ g/cm$^{3}$),
the exact value depends on the equation of state) we used a three-segment
approximant, fitted to the equation of states listed above. For low-baryonic
mass densities ($\rho < \rho_{\rm core}$), matter is described using a
four-segment approximant fitted against the SLy equation of
state~\cite{Read:2008iy,Kumar:2019xgp}.

One could argue~\cite{Miller:2019nzo} that a more rigorous approach to do
inference on the nuclear equation of state from neutron star observations is to
not use hard cuts as we do here. However, since our approach makes use of
equation-of-state independent relations, changes to our equation of state
catalog imply only small changes to the numerical values of the fitting
coefficients. The hard cuts then have a negligible impact on our inferences,
since the uncertainty on our quoted results (e.g., in Table~\ref{tab:gr_infer})
are dominated by the uncertainties on the observations, except for the
eccentricity $e_{\star}$.

\bfi{Neutron-star catalog.}
%
For each equation of state, we calculate a
sequence of 50 neutron stars, spanning central baryonic mass densities in the
range
$\rho_{\rm c} \in [1.5, 8.0] \, \rho_{\rm nuc}$
(where $\rho_{\rm nuc} = 2.8 \times 10^{14}$ g/cm$^{3}$ is the nuclear
saturation density)
in general relativity and in dynamical Chern-Simons gravity.
To obtain neutron star solutions in dynamical Chern-Simons gravity we use the
same perturbative expansion in small-spin ($\varepsilon \ll 1$) as used in
general relativity, outlined in the main text, but combined with a
small-coupling ($\zeta \ll 1$) approximation as required for the well-posedness
of the theory~\cite{Yagi:2013mbt}.
We then retained only those with
(i) $M \leqslant M_{\rm max}$ (i.e.~linearly stable against radial
perturbations) and
(ii) with $\cC \in [0.125, 0.3125]$ (i.e.~within the prior on $\cC$ used in the
parameter estimation of \idd by the Illinois-Maryland MCMC samples~\cite{miller_m_c_2019_3473466}).
In the case of dynamical Chern-Simons gravity only we further imposed (iii)
that all stars have $\zeta \leqslant 0.2$ (the small-coupling approximation).
Although we used a $\cC$-interval in (ii) based on~\cite{miller_m_c_2019_3473466} we
note that 97\% of the MCMC samples of the analysis by Riley et al.~\cite{Riley:2019yda}
do fall within this range.
Therefore, we use the same $\cC$ relations to make inferences with both MCMC
samples, instead of deriving two sets of $\cC$ relations based on the two NICER
analysis~\cite{Miller:2019cac,Riley:2019yda}.

\begin{figure*}[th!]
\includegraphics[width=\columnwidth]{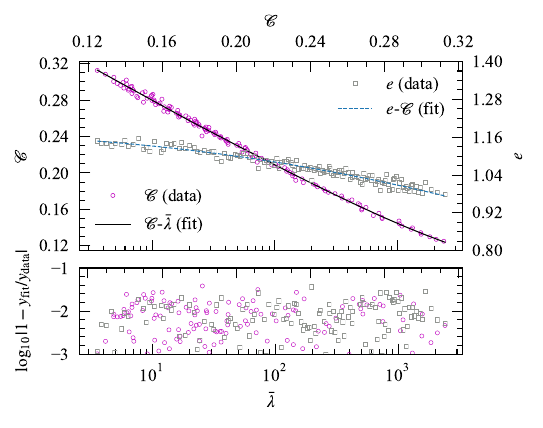}
\includegraphics[width=\columnwidth]{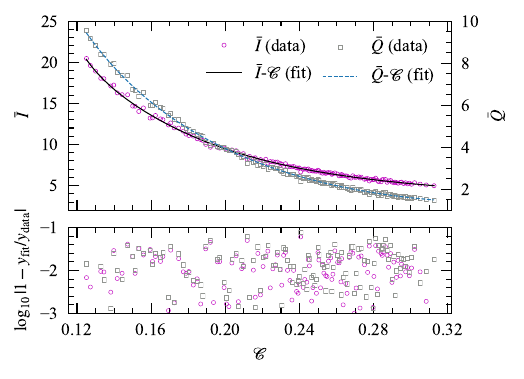}
\caption{The $\cC$ relations.
The top-left panel shows the relation between tidal Love number $\bl$ and
compactness $\cC$ (bottom and left axes) and between eccentricity $e$ and
compactness $\cC$ (top and right axes).
The top-right panel shows the relation between the moment of inertia $\bi$
(left axes), rotational quadrupole moment $\bq$ (right axes) and compactness
$\cC$.
The lower panels show the relative errors
($\epsilon = |1 - y_{\rm fit} / y_{\rm data}|$)
between the data and the fits. The largest mean relative error among the four
$\cC$ relations is $2\%$ (corresponding to the Q-$\cC$ relation), which shows
the equation-of-state insensitivity of the relations.
}
\label{fig:c-relations}
\end{figure*}

\bfi{The $\cC$ relations.}
%
Relations among some stellar observables (e.g., moment of inertia, Love number or quadrupole moment) that do not depend sensitively on the equation-of-state are called quasi-universal relations~\cite{Yagi:2016bkt}. These relations are of the form
\begin{equation}
y = \sum_{i = 0}^{n} a_i x^{i}\,,
\label{eq:fit}
\end{equation}
where $y$ and $x$ are particular stellar observables for which the relation exists, and $a_{i}$ are constant coefficients.
In this work we use the ``$\cC$ relations''~\cite{Lattimer:2015nhk,Maselli:2013mva,Baubock:2013gna,Breu:2016ufb}, shown in Fig.~\ref{fig:c-relations}, with
$(y, \, x) = \{
(\log_{10}\bi_{\GR}, \, \cC^{-1}),
\, (\cC, \, \log_{10}\bar{\lambda}_{\GR}),
\, (\log_{10}\bar{Q}_{\GR}, \, \cC),
\, (e, \, \cC)
\}$.
The values of the coefficients $a_{i}$, the number of coefficients $n$, and the
mean $\left\langle \epsilon^{y} \right\rangle $ and maximum relative error
$\epsilon^{y}_{\rm max} = \max|1 - y / y_{\rm fit}|$ of the resulting fits
are presented in Table~\ref{tab:coeff}.
In comparison to the I-Love relation, the $\cC$ relations are slightly
``less'' quasi-universal (10\% versus 1\% maximum relative error).
However, their typical mean relative error at the 1\% percent level justifies their
use for parameter inference.
We also verified explicitly that the $\cC$ relations are
weakly dependent on our equation of state catalog.
This was verified by comparing the resulting fits (and associated errors)
obtained using the initial full set of 85 equations and its subset of 46.

\begin{table*}[htpb]
\begin{tabular}{ c c c c c c c c c c }
   \hline
   \hline
   $y$ & $x$ & $n$ & $a_0$ & $a_1$ & $a_2$ & $a_3$ & $a_4$ & $\left\langle \epsilon^{y} \right\rangle $ & $\epsilon^{y}_{\rm max}$\\
   \hline
   $\log_{10}\bar{I}_{\GR}$ & $\cC^{-1}$ & 4 & $5.5531\times 10^{-1}$ & $-1.7705\times 10^{-1}$ & $1.1105\times 10^{-1}$ & $-1.5137\times 10^{-2}$ & $6.9401\times 10^{-4}$ & $1.52\times 10^{-2}$ & $6.7\times 10^{-2}$\\
   $\cC$ & $\log_{10}\bar{\lambda}_{\GR}$ & 4 & $3.5818\times 10^{-1}$ & $-8.8149\times 10^{-2}$ & $1.3120\times 10^{-2}$ & $-4.5810\times 10^{-3}$ & $6.9924\times 10^{-4}$ & $9.63\times 10^{-3} $& $4.2\times 10^{-2}$\\
   $\log_{10}\bar{Q}_{\GR}$ & $\cC$ & 4 & $2.1088$ & $-1.4669\times 10^{1}$ & $6.6952\times 10^{1}$ & $-2.1050\times 10^{2}$ & $2.6576\times 10^{2}$ & $1.91\times 10^{-2}$& $8.6\times 10^{-2}$\\
   $e$ & $\cC$ & 2 & $1.1624$ & $ 2.0527\times 10^{-1}  $ & $-2.5870$ & - & -  & $9.86\times 10^{-3}$& $4.1\times 10^{-2}$\\
   \hline
   \hline
\end{tabular}
\caption{Numerical coefficients for the fits to the $\cC$ relations.
These fits are tailored to the Illinois-Maryland MCMC
data samples~\cite{miller_m_c_2019_3473466} and, therefore, are only valid within
the compactness range $\cC \in [0.125, 0.3125]$.}
\label{tab:coeff}
\end{table*}

\bfi{Parameter inference: \id.}
%
To calculate the posterior distributions
for $y = \{ \bi$, $\bl$, $\bar{Q}$, $e\}$, we follow the procedure used
in~\cite{Kumar:2019xgp}.
In this scheme, the maximum relative error $\epsilon_{\textrm{max}}^{y}$ (cf.
the last column of Table~\ref{tab:coeff}) is used to define the half-width of
the 90\% credible interval of a Gaussian distribution centered at each fitted
value ($y_{\rm fit}$),
\begin{equation}
P\left(y|\cC \right)=(2\pi\sigma_{y}^{2})^{-1/2}
\exp\left[-(y-y_{\rm fit})^{2} / (2\sigma_{y}^{2}) \right]\,,
\label{eq:GauError}
\end{equation}
where $\sigma_{y}=\epsilon_{{\rm max}}^{y} \, y_{{\rm fit}}/1.645$.

\begin{figure}[t]
\hspace*{-.5cm}\includegraphics[width=\columnwidth]{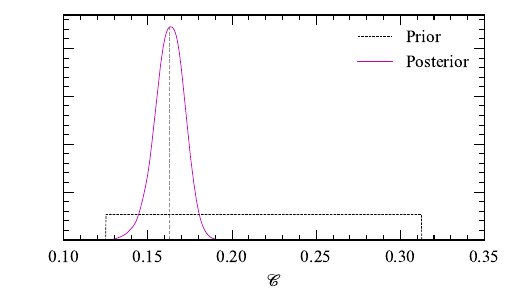}
\caption{Prior and posterior probability distributions on the compactness
of \id\, ($\cC_{\star}$).
The Illinois-Maryland analysis~\cite{Miller:2019cac,miller_m_c_2019_3473466}
assumed a flat prior probability distribution for $\cC_{\star}$ with bounds
$0.125$ and $0.3125$ (dashed curve). The posterior probability distribution
(solid line) has a median value of $\cC_{\star} =
0.163$~\cite{Miller:2019cac}.
The posterior is approximately Gaussian and very different from the flat
(uninformative) prior used by NICER, showing that the NICER observation was
indeed informative.
}
\label{fig:cpp}
\end{figure}

The posterior distribution for each dimensionless quantity,
$P\left(y|{\rm NICER}\right)$,
is then calculated using the corresponding $\cC$ relation and
the posterior probability distribution of the compactness,
$P\left(\cC|{\rm NICER}\right)$
(see Fig.~\ref{fig:cpp}), obtained directly using NICER's MCMC
samples~\cite{miller_m_c_2019_3473466,riley_thomas_e_2020_3707821}, and then
marginalizing over $\cC$:
\begin{equation}
P\left(y|{\rm NICER}\right) =
\int P\left(y|\cC \right)P\left(\cC|{\rm NICER}\right)\, \dd\cC\,.
\label{eq:posterior}
\end{equation}

\begin{figure}[t]
\includegraphics[width=\columnwidth]{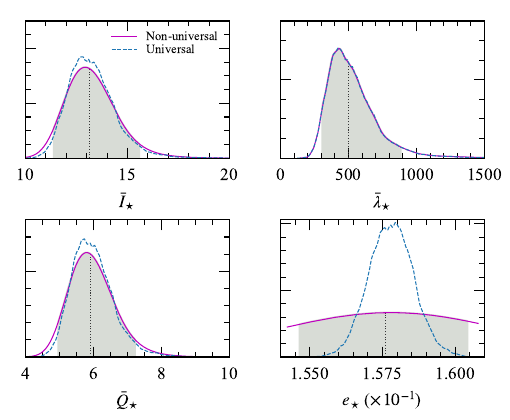}
\caption{Posterior distributions on the dimensionless quantities
$\{ \bi$, $\bl$, $\bar{Q}$, $e$ $\}$ of \id.
We show both the distributions obtained by taking into account the systematic
error introduced by the quasiuniversality of the $\cC$ relations (solid
lines) and that assuming the universality of the relations (dashed lines).
The vertical lines and the shaded bands represent the mean and 90\% posterior
credible intervals, respectively.
The posteriors show that for $\bi$, $\bl$ and $\bq$, the uncertainties
associated with our inferences are dominated by the uncertainty on the measured
value $\cC_{\star}$. This is not the case for $e$, because the systematic error
associated with the e-$\cC$ fit is larger than that associated with the inference
of $e$, assuming complete equation-of-state independence on the e-$\cC$ relation.}
\label{fig:psdim}
\end{figure}

What dominates the uncertainty on our inferred values for the neutron star
parameters? Is it the uncertainties associated with the $\cC$-measurement
we are using, or those introduced by the approximate universality of the
$\cC$ relations?
To answer this question, in Figure~\ref{fig:psdim} we show the resulting
posterior distributions by taking into account the approximate universality
of the $\cC$ relations (solid lines) and compare against the posterior
distributions on the same parameters if we do not (dashed lines).
In the latter case, the posteriors are obtained by directly applying
the $M$-$R_{\rm e}$ samples~\cite{miller_m_c_2019_3473466} in the
$\cC$ relations.
The effect of including the systematic error induced by the variability of the
$\cC$ relations is to broaden the posterior, a small effect for $\bi$, $\bl$
and $\bq$, but not for the eccentricity $e$.
The reason is the following: the mean relative error for the
e-$\cC$ fit is approximately 4\%, which is larger than the variability of $e$ in
the 1$\sigma$ interval of its distribution. This leads to the considerable
broadening shown in the bottom-right panel of Fig.~\ref{fig:psdim}.

The posteriors for the dimensionful version of these
quantities are obtained of through a change of variables, and using the
posteriors on $M$ and $R_{\rm e}$. For instance, for the moment of inertia $I$,
\begin{align}
P(I | \textrm{NICER}) &= (c^2 / G)^2 \int P(c^4 I / (G^2 M^3) | \textrm{NICER})
\nonumber \\
                      &\quad\times P(M | \textrm{NICER})\, M^{-3} \, \dd M\,.
\end{align}
The posterior distributions for the dimensionful variables are similar in
shape to their nondimensionalized versions.


\bfi{Parameter inference: fixed mass.}
%
To calculate the posterior distribution of a quantity $y$ for a neutron star
with known mass $M_0$ using the $\cC$ relations, we first need to contruct a
posterior
$P(\cC_{M_0} | {\rm NICER})$
given
$P(\cC | {\rm NICER})$.
To do so, we write
\begin{equation}
P(\cC_{M_0} | {\rm NICER}) = P( [G M / (R_{\rm e} c^2)]_{M = M_0} | {\rm NICER})\,,
\end{equation}
where the right-hand side is obtained by using the MCMC
samples~\cite{miller_m_c_2019_3473466}, doing the substitution
$\{M, R_{\rm e}\} \mapsto \{M_0, R_{\rm e}\}$
at each sample and then computing $\cC_{M_0}$.
This procedure gives an approximation to the distribution of
compactness for a system with known mass $M_0$, which we here take to be identical to \id.
Of course, the posterior $\cC_{M_0}$ obtained in this way would fail dramatically
if the difference between $M_0$ and \id's mass, $M_{\star} = 1.44^{+0.15}_{-0.14}\,\msun$~\cite{Miller:2019cac},
is large.
However, this is not case for a canonical neutron star with mass $1.4 \,\msun$
and \iddp's, which has a mass of $1.3381\,\msun$ masses, both of which agree
with $M_{\star}$ within 1$\sigma$.

\begin{figure}[t]
\includegraphics[width=0.85\columnwidth]{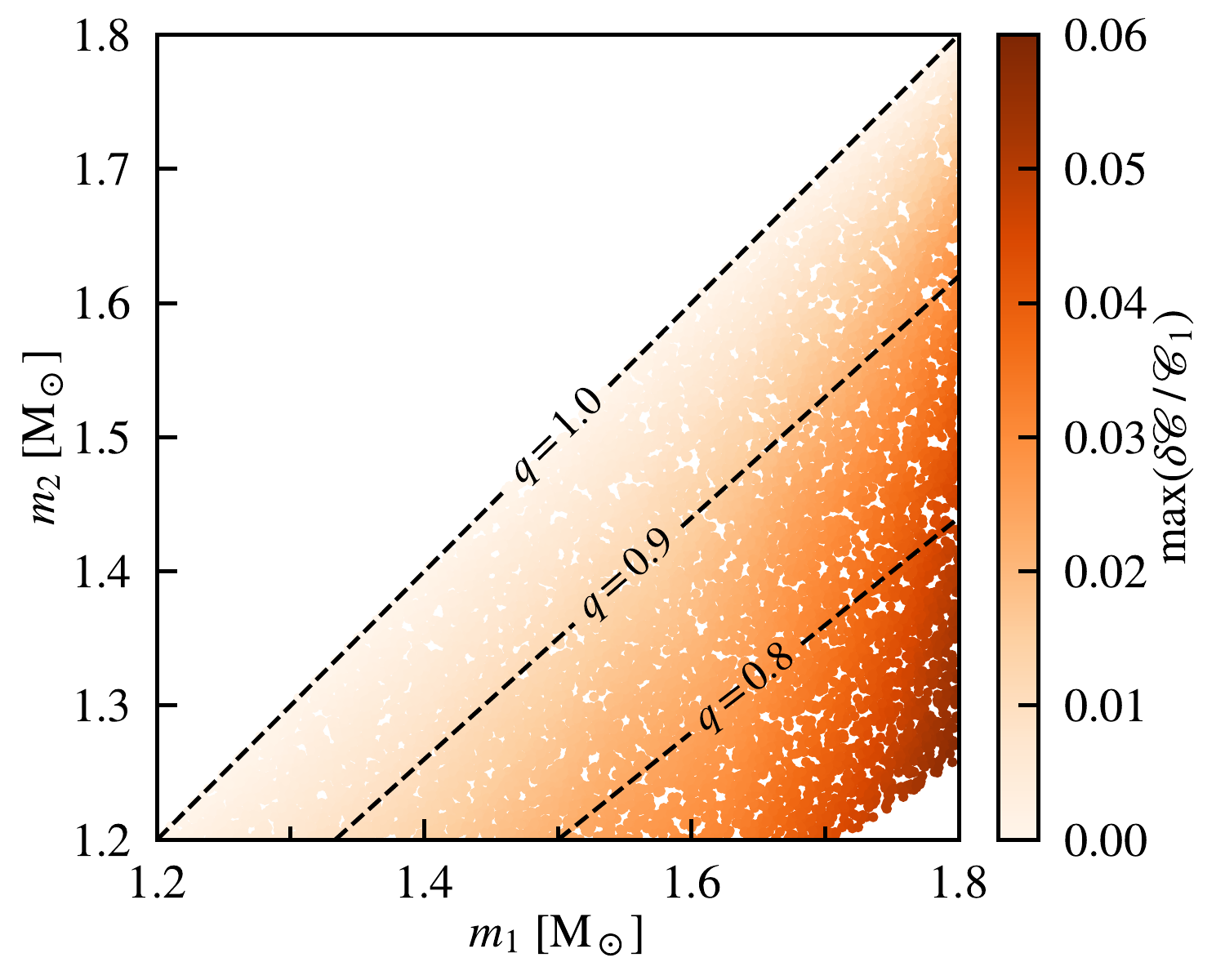}
\caption{Maximum residual when predicting the compactness $\cC_2$ of a NS with
$m_2$ given $m_1$ and $\cC_1$. The dashed contours correspond to lines of
constant mass ratio $q = m_2 / m_1$. The mass ratio between PSR~J0030+0451 and
both a PSR~J0737-3039A and a $1.4~\msun$ star is about $0.9$.
Overall, the largest residual is about $0.06$ in the mass range shown.}
\label{fig:extrapolation}
\end{figure}

We can justify this more quantitatively as follows.
First, for each equation of state we compute the compactness
for each mass $m_1$ and $m_2$. For our case, this would produce 46 points in
the $\cC_1$ and $\cC_2$ plane, corresponding to the 46 equations of state we use.
Then, we fit a line to these points and calculate the maximum residual from the
best-fit line.
This procedure can be done for any combinations of $m_1$ and $m_2$.  We plot
the maximum residual for the estimated compactness for different $m_1$ and
$m_2$ in Fig.~\ref{fig:extrapolation}.
The mass ratio between PSR J0030+0451 and both PSR J0737-3039A and a $1.4~\msun$ star
is approximately $0.9$ for which the maximum residual is about 1\%.

Knowing $P(\cC_{M_0} | {\rm NICER})$, we can use it in Eq.~\eqref{eq:posterior}
and marginalize over $\cC_{M_0}$ to obtain the posterior distribution of a
quantity $y$ at mass $M_0$.
Our results for $\cC$, $R_{\rm e}$ and $y = \{\bi$, $\bl$, $\bar{Q}$, $e\}$
for a canonical neutron star are summarized in Table~\ref{tab:gr_infer14}.

\begin{table*}[htpb]
\begin{tabular}{ l c c c c }
\hline
\hline
Parameter & Source / Universal relation & Median & $-1.65\sigma$ & $+1.65\sigma$ \\
\hline
$\cC_{1.4}$ (10$^{-1}$) & MCMC chain / $-$ & 1.59 & 1.37 & 1.84 \\
$R_{1.4}$ (10) [km] & MCMC chain / $-$ & 1.30 & 1.12 & 1.51 \\
\hline
$\bar{I}_{1.4}$  ($10$)  & MCMC chain / I-$\cC$ & 1.46 & 1.12  & 1.91 \\
$\bar{\lambda}_{1.4}$ ($10^2$) & \quad\,\, MCMC chain / Love-$\cC$ & 5.80 & 2.33 & 13.7 \\
$\bar{Q}_{1.4}$ & \,\,MCMC chain / Q-$\cC$ & 6.12 & 4.52   & 8.12   \\
$e_{1.4,\, 0}$   & \hspace{-.005cm}MCMC chain / e-$\cC$ &  1.12 &  1.09 & 1.15   \\
\hline
\hline
\end{tabular}
\caption{Properties of $M = 1.4\,\msun$ neutron stars, inferred from \idd at
90\% credibility. The eccentricity $e_{1.4,\,0}$ is evaluated at the characteristic
mass-shedding frequency $f_0$. To rescaled it to the eccentricty $e_{1.4}$ of a
star spinning with frequency $f$ one has to multiply by $\varepsilon =f / f_0 =  1532.7$ Hz.}
\label{tab:gr_infer14}
\end{table*}

\bfi{I-Love relation in dynamical Chern-Simon gravity.}
%
\begin{figure}[t]
\includegraphics[width=\columnwidth]{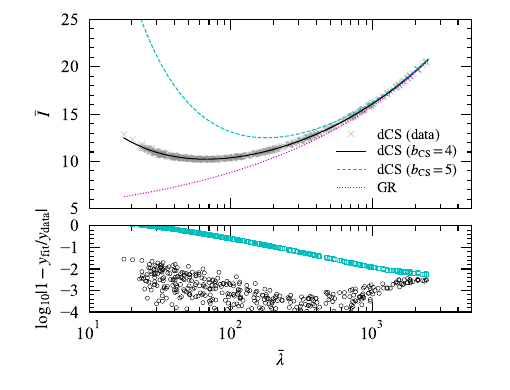}
\caption{I-Love relation in dynamical Chern-Simons gravity.
In the top-panel we show the numerical data (crosses) obtained for neutron
stars in dynamical Chern-Simons, the I-Love relation in general relativity
(dotted line) and fits to it using the Newtonian result obtained in the
main text (dashed line) and the post-Minkowskian corrected result (solid line).
In the bottom-panel we show the relative errors between both fits and the data,
explicitly showing that the Newtonian fit cannot be used for an I-Love test.
In this figure we fixed $\bx = 10^{3}$ and only show stars with $\zeta \leqslant 0.25$.
}
\label{fig:ilcs}
\end{figure}
Let us describe further how
to map the theory-agnostic deformation of the I-Love relations in
Eq.~\eqref{eq:ilove_param} to dynamical Chern-Simons gravity.
The procedure outlined here can be applied to other theories as well.

The dimensionless moment of inertia of a constant density star in dynamical Chern-Simons gravity is
$\bi= \bi_{\GR} + \bi_{\CSN}$, where $\bi_{\CSN} = C_{\bi\bl}^{\CSN} \alpha^2 M / R^5$
and $C_{\bi\bl}^{\CSN} = 1024 \pi/75$~\cite{AliHaimoud:2011fw}
to leading-order in $\alpha$ and leading-order in $\cC$.
The I-Love relations in dynamical Chern-Simons gravity are insensitive
to the equation of state when $\bi_{\CS}$ is normalized via $\bx \equiv 16 \pi \alpha^{2} / M^4$~\cite{Gupta:2017vsl}.
Therefore, using that $\bl_{\N} = (1/2) \cC^{-5}$ in Newtonian gravity for a constant density star,
we then have that $\bi_{\CSN} = (16 \pi)^{-1} \, C_{\bi\bl}^{\CSN} \, \bx \, \cC^{5}  = [(32 \pi)^{-1} \, C_{\bi\bl}^{\CSN} \, \bx] \, \bl^{-1}$.
Although this relation was derived assuming a constant density star, it holds for any equation of state,
with variability at the sub-percent level~\cite{Gupta:2017vsl}.
Comparison of this result to Eq.~\eqref{eq:ilove_param} reveals that the
mapping between our proposed parameterization and dynamical Chern-Simons
gravity is $\beta_{\CSN} = (32/75) \, \bx$ and $b_{\CSN} = 5$.

How well does this ``Newtonian'' approximation capture the fully relativistic
I-Love relation in dynamical Chern-Simons gravity?
Figure~\ref{fig:ilcs} shows that the Newtonian approximation derived above
is excellent for low-mass neutron stars.  The scaling with $\bl^{-1}$, however,
fails for more massive stars, when
$M \gtrsim 1\,\msun$,
because relativistic corrections become important.
In spite of this, the dynamical Chern-Simons correction to the relativistic
I-Love data, for values of $\bl$ corresponding to stars with compactness
$\cC$ within NICER's priors, can be well-approximated by
$\bi_{\CS} = \beta_{\CS}  \, \bl^{-b_{\CS}/5}$, with
\begin{equation}
\beta_{\CS} = 6.15 \times 10^{-2} \, \bx\,,\quad b_{\CS} = 4\,.
\end{equation}
This functional form is quite close to the Newtonian result (in fact, just a
factor of $\cC$ higher than the Newtonian result because $\cC \propto
\bl^{-1/5}$), which suggests the rule of thumb $b = b^{\N} - 1$, where $b^{\N}$
is the result of the I-Love calculation to Newtonian order in modified gravity.

By virtue of the parity properties of the field equations in dynamical
Chern-Simons gravity, the electric-type $\ell=2$ tidal Love number in
this theory is identical to that in general relativity for nonrotating
neutron stars.

Using the results from Riley et al.~\cite{Riley:2019yda,riley_thomas_e_2020_3707821},
we obtain the same bound on dCS found in the main text, namely $\sqrt{\alpha} \leqslant 8.5$~km.
We also find that $\zeta \leqslant 0.26$, only 4\% above our threshold for the small-coupling approximation,
however still $\ll 1$.

\bfi{Constraints on the parametrized I-Love parameters}
%
In Table~\ref{tab:ilovepm} we summarize the numerical values for the contrainst
of the parametrized I-Love parameters $b$ and $\beta$.
\begin{table*}[htpb]
\begin{tabular}{ c | c c c c c c c c}
   \hline
   \hline
   $b$ & $-2$ & $-1$ & 0 & 1 & 2 & 3 & 4 & 5 \\
   \hline
   $\beta_{\crit}$ (Miller et al.) & 1.97 & 4.63 & 10.8 & 25.7 & 74.5 & $2.49 \times 10^{2}$ & $8.84 \times 10^{2}$ & $ 3.15\times 10^{3}$ \\
   $\beta_{\crit}$ (Riley et al.)  & 1.99 & 4.67 & 11.0 & 26.0 & 75.5 & $2.53 \times 10^{2}$ & $9.00 \times 10^{2}$ & $ 3.21\times 10^{3}$ \\
   \hline
   \hline
\end{tabular}
\caption{\label{tab:ilovepm} Constraints on the parametrized I-Love relation.
For a given exponent $b$, parametrized I-Love curves with $\beta \geqslant
\beta_{\crit}$ do not pass through the 90\% credible intervals on the inferred
values of $\bi_{1.4}$ and $\bl_{1.4}$ as shown in Fig.~\ref{fig:pilove}. The constraints obtained
using the results of both Miller et al.~\cite{Miller:2019cac} and Riley et al.~\cite{Riley:2019yda}
are similar.
}
\end{table*}

\bfi{The parametrized I-Love test and why can neutron stars constrain dynamical Chern-Simons gravity}
%
Here we explain the rationale behind the parametrized I-Love
test of gravity and why it can be applied to dynamical Chern-Simons gravity.
Perhaps the easiest way to understand the parametrized I-Love test is to
present it alongside with the parametrized post-Keplerian formalism (ppK) used to test
general relativity with binary pulsars~\cite{Damour:1991rd}. Both are very similar in spirit.
In the ppK formalism, a timing model that assumes Keplerian orbits augmented with
certain theory-agnostic post-Keplerian parameters ${\bm \theta}$ is fit to timing data. This
provides marginalized posterior distributions for all parameters in the timing
model, including both the orbital ones (such as the orbital period $P_b$ and
eccentricity $e$), as well as the post-Keplerian ones (such as
the Shapiro time delay $\gamma$ and the secular advance of periastron $\dot{\omega}$).
To do a theory-agnostic, general relativity test with binary-pulsar
data, one then uses expressions that assume general relativity is correct for
how the post-Keplerian parameters depend on the component masses $m_A$ and
$m_B$.
These expressions then define curves in $m_A$--$m_B$ space (one curve per
post-Keplerian parameter), and under the assumption that general relativity is
correct, then all curves must interest at the same point in $m_A$--$m_B$ space
(see e.g.~\cite{Kramer:2006nb}).
In practice, the curves have a width since the measurements are not perfect, so
the intersections of the curves define a two-dimensional region in $m_A$--$m_B$
space.

Let us provide a concrete example. Assuming general relativity, the post-Keplerian parameters
$\dot{\omega}$, $\gamma$ and $r$ (the range of the Shapiro time delay) are
%
$\dot{\omega} = {3 n}{(1 - e^2)^{-1}} \left( {G M n}/{c^{3}} \right)^{2/3}$,
$\gamma = (e/n) X_B \left( {G M n}/{c^{3}} \right)^{2/3} \left(1 + X_B\right)$ and
$ r = {G m_B}/{c^{3}}$,
where $X_{A,B} = m_{A,B}/M$ is the mass ratio, with the total mass $M=m_A +
m_B$ and component masses $m_A$ and $m_B$, while $e$ is the orbital
eccentricity and $n = 2 \pi/P_b$ is the orbital frequency with $P_b$ the
orbital period.
Since the orbital period and the eccentricity are independent
parameters in the timing model that are, therefore, included in the Bayesian
parameter estimation, the above expressions represent 3 curves in $m_A$--$m_B$
space. The curves have a width that is dominated by the uncertainty in the
measurements of $\gamma$, $\dot{\omega}$ and $r$, so if the assumption that
general relativity is correct, these three finite-width curves must intersect
and define a two-dimensional region in $m_A$--$m_B$ space.

Our theory-agnostic test is very similar to this one.
In our approach, the quantities that are \emph{independently} measured are the
moment of inertia $I$ and the Love number $\lambda$, which require
two independent observations (NICER and LIGO/Virgo ones, instead of
a single binary pulsar observation).
These two measurements define a two-dimensional region in $I$--$\lambda$ space,
whose finite size is associated with the uncertainty in these observations.
The moment of inertia and the Love number also satisfy a relation, the I-Love
relation, which is (approximately) independent of the equation of state, but
dependent on the theory of gravity. Therefore, the relation constructed
assuming general relativity must cross the two-dimensional box in $I$--$\lambda$
space defined by the two observations if general relativity is correct.
This does indeed happen as we showed for the first time in Fig.~\ref{fig:pilove}
of the main text.

How do these tests work to constrain a specific modified-gravity theory? In the
binary pulsar case, the timing model used to analyze the data does not change:
it is still the Keplerian model enhanced with post-Keplerian parameters ${\bm\theta}$.
What changes is the relation between these post-Keplerian parameters and the
component masses, a relation that will now also typically include additional
modified theory parameters.
For example, in scalar-tensor theories~\cite{Damour:1992we} that support
spontaneous scalarization~\cite{Damour:1993hw}, this relation now also depends
on the spontaneous scalarization parameter $\beta$~(see
e.g.~\cite{Damour:1996ke,Anderson:2019eay}).
Therefore, the observation of three post-Keplerian parameters allows one to
infer a constraint on $\beta$, since if this parameter were large (and
negative) enough to generate a strong deviation from general relativity, then
the three post-Keplerian parameter curves would not intersect in the
$m_1$--$m_2$ space.

Our constraint on specific modified-gravity theories operates in a similar way,
and just for the sake of concreteness, let us consider dynamical Chern-Simons
gravity as a simple example.
We will assume that the model used to fit the data (i.e.~the lightcurve model
used by NICER and the gravitational waveform model used by LIGO/Virgo)
is that assumed in general relativity. This leads to a two-dimensional region in
$I$-$\lambda$ space, just as in the theory-agnostic case.
What changes now is that the $I$--$\lambda$ relation is not just a function of
the Love number, but it also depends on the coupling parameter $\alpha$.
The larger this parameter is, the more the $I$--$\lambda$ curve will deviate
from the observed two-dimensional region in $I$-$\lambda$ space.
One can then determine how small $\alpha$ must be in order to be consistent
with the $I$ and $\lambda$ observations, and this is the constraint quoted
in the main text. We summarize this discussion in Table~\ref{tab:ppk_pil_comp}.

\begin{table*}[t!]
\begin{tabular}{ l | c | c }
    \hline
    \hline
                                & ppK                                                   & pIL \\
    \hline
    Astrophysical system        & binary pulsar                                         & isolated NS and a NS binary        \\
    \hline
    Instrument                  & radio telescopes                                      & NICER and LIGO/Virgo      \\
    \hline
    Model                       & theory-agnostic timing formula                        & GR light-curve and GW waveform    \\
    \hline
    Parameters                  & post-Keplerian ${\bm \theta}(m_1, m_2, \mathscr{P})$        & $I = I(\lambda, \mathscr{P})$           \\
    \hline
    \# of parameter                & 3 (two for $m_1$, $m_2$,                              & 2                                 \\
    measurements for test       & third for test)                                       &                                   \\
    \hline
    \hline
\end{tabular}
\caption{Comparison between the parametrized post-Keplerian (ppK) and parametrized I-Love (pIL) relation tests of gravity. Here $\mathscr{P}$ collectively represents the parameters that control the deviations from general relativity. They all vanish for general relativity.}
\label{tab:ppk_pil_comp}
\end{table*}

Is our assumption that the model used to fit the NICER and LIGO/Virgo data is
not modified from general relativity a good assumption to make in this test?
Let us first continue considering dynamical Chern-Simons gravity.
In this theory it has been shown that neutron stars do not have (monopole)
scalar hair~\cite{Wagle:2018tyk}.
Therefore one cannot use the gravitational waves emitted during the inspiral
of binary neutron stars to test dynamical Chern-Simons gravity due to the
absence of dipole scalar radiation emission.
Moreover, the light-curve model used by NICER considers the emission of X-rays
from an oblate star surface and with photons propagating in a Schwarzschild
spacetime (the so-called ``Oblate+Schwarscshild model''~\cite{Morsink:2007tv}).
On the one hand, non-rotating stars in dynamical Chern-Simons gravity are
identical to those of general relativity, and therefore, their exterior
spacetime is also described by the Schwarzschild metric~\cite{Yunes:2009ch}.
On the other hand, the stellar oblateness (due to rotation) could in principle
be used to distinguish this theory gravity from general relativity.
However modifications to the oblateness due to changing the theory of gravity
are degenerate with assuming general relativity but changing the underlying
equation of state.
Together, these points justify our assumption that the model used to fit the
NICER and LIGO/Virgo data is not modified from general relativity in dynamical
Chern-Simons gravity.

What about other theories? In the case of scalar-tensor theories, which support
spontaneous scalarization (exemplified above), neutron stars have (monopole)
scalar hair and therefore, in general, emit dipolar scalar radiation when found
in binaries; and they also have an exterior spacetime that differs from the
Schwarzschild metric. In this case, a direct confrontation of the models in
this theory against observational data could yield a constraint on $\beta$ using
observations by NICER (see~\cite{Sotani:2017rrt,Silva:2018yxz,Xu:2020vbs} and
particularly~\cite{Silva:2019leq} for initial work in this direction).
Such constraints are likely to be stronger than the ones obtained using our
parametrized I-Love test.
This is specially true for theories whose lightcurve model and gravitational
waveforms are considerably different from general relativity.
But none of this invalidates our I-Love test. At the very least, our I-Love
test would provide complementary constraints to other more direct tests.
More importantly, the I-Love test is a new tool to constrain theories of
gravity \emph{which would remain unconstrained by NICER and gravitational wave
observations alone}, as we have shown for dynamical Chern-Simons
gravity as an example.
This fact highlights the importance of using multimessenger observations to
constrain deviations from general relativity.
Our parametrized I-Love test is the first one in this direction.

Let us do a back-of-the-envelope estimate to explain \textit{why} neutron star
observations have such a constraining power in dynamical Chern-Simons gravity.
Assume that $\zeta$ has been constrained to some value, say $\zeta \leqslant
0.25$, by some set of observations that involve also measurements of the mass
and equatorial radius of a neutron star.
A posterior distribution for ${\alpha}^{1/2}$ can then be obtained by
translating the MCMC samples for $M$ and $R_{\rm e}$ into samples of
${\alpha}^{1/2} = [\zeta / (16 \pi)]^{1/4} \cR$.
Using for concreteness the MCMC samples obtained by NICER, one would obtain
${\alpha}^{1/2} \leqslant 8.57^{+0.93}_{-1.13}$ km (at 90\% credibility), which
is very close to the real constraint derived in the main text.
The back-of-the-envelope estimate presented above is of course
\emph{not an actual constraint since NICER alone cannot bound $\zeta$
due to degeneracies with the equation of state}.
But this estimate does tell us that such a test would be dominated by $\cR$,
because an improvement on the bound on $\zeta$ by a factor of $k$ would only
improve the constraint on ${\alpha}^{1/2}$ by $k^{1/4}$.

\end{document}